
\input phyzzx
\overfullrule=0pt
\def\ETslash{\not{\hbox{\kern-4pt $E_T$}}}

\def\ra{\rightarrow}

\def\WW {W_L W_L}
\def\MWW {M_{WW} }
\def\WWWW {W_L W_L \ra W_L W_L}
\def\PPPP {\phi \phi \ra \phi \phi}
\def\WPWM{ W^+(\ra l^+\nu) W^-(\ra q_1 \bar q_2) }
\def\WPZ{ W^+(\ra l^+ \nu) Z^0(\ra q \bar q) }
\def\WPWP{ W^+(\ra l^+ \nu) W^+(\ra l^+ \nu) }

\nopubblock
\line{\hfil MSUTH 92/06 }
\line{\hfil November, 1992}

\titlepage
\title{ Proposals for Studying TeV $\WWWW$ Interactions Experimentally\foot{
To be published in ``Perspectives On Higgs Physics,''
edited by Gordon L. Kane, World Scientific, Singapore, in press.} }
\author{ C.--P. Yuan}
\address{
Department of Physics and Astronomy \break
Michigan State University  \break
East Lansing, MI 48824}

\abstract{
We discuss how to experimentally study the symmetry
breaking sector by observing
$\WWWW$ interactions in the TeV region.
We discuss some general features of the event structure in the
signal and background events. Various tricks
to enhance the signal--to--background ratio
are also presented. We show how to detect longitudinal $W$--bosons
either in the central rapidity region of the detector or in
the beam pipe direction.
}

\endpage

\chapter{Introduction}

\REF\effwone{
R. N. Cahn and S. Dawson, Phys. Lett. {\bf B136}, 196 (1984), Phys.
Lett. {\bf B138}, 464(E) (1984).}
\REF\effwtwo{
M. S. Chanowitz and M. K. Gaillard, Phys. Lett. {\bf B142}, 85 (1984);
G. L. Kane, W. W. Repko and W. R. Rolnick, Phys. Lett. {\bf B148}, 367 (1984);
S. Dawson, Nucl. Phys. {\bf B249}, 42 (1985).}
\REF\effwthree{J. Lindfors, Z. Phys. {\bf C28}, 427 (1985);
W. B. Rolnick, Nucl. Phys. {\bf B274}, 171 (1986); P. W. Johnson, F. I. Olness
and Wu--Ki Tung, Phys. Rev. {\bf D36}, 291 (1987);
Z. Kunszt and D. E. Soper, Nucl. Phys. {\bf B296}, 253 (1988); A. Abbasabadi,
W. W. Repko, D. A. Dicus and R. Vega, Phys. Rev. {\bf D38}, 2770 (1988); S.
Dawson, Phys. Lett. {\bf B217}, 347 (1989).}
\REF\ETone{
J. M. Cornwall, D. N. Levin, and G. Tiktopoulos,
Phys. Rev. {\bf D10}, 1145 (1974); C.~Vayonakis, Lett. Nuovo Cimento {\bf 17},
383 (1976); B.~W.~Lee, {\hbox {C. Quigg}}, and H.~Thacker,
Phys. Rev. {\bf D16}, 1519 (1977).}
\REF\ETtwo{
M.~S.~Chanowitz and M.~K.~Gaillard, Nucl. Phys. {\bf B261}, 379 (1985);
G.~J.~Gounaris, R.~Kogerler, and H.~Neufeld, Phys. Rev. {\bf D34}, 3257
(1986).}
\REF\ETthree{
Y.--P. Yao and C.--P. Yuan, Phys. Rev. {\bf D38}, 2237 (1988);
J. Bagger and C. R. Schmidt, Phys. Rev. {\bf D41}, 264 (1990);
H. Veltman, Phys. Rev. {\bf D41}, 2294 (1990);
H.--J. He, Y.--P. Kuang and Xiaoyuan Li, CCAST preprint, 1992.}

The Standard Electroweak Model has been tested and is
very successful
in explaining and predicting experimental data. However, we still do
not have any understanding about the origin of the fermion masses or
the spontaneous
symmetry breaking mechanism (Higgs mechanism) of the Standard Model (SM).
To probe the symmetry breaking sector, we need to detect the longitudinal
$W$ pairs produced via the $\WW$ fusion
mechanism\refmark{\effwone - \effwthree}
if a light Standard Model Higgs boson does not exist.
(We shall use $W$ to denote either $W^\pm$ or $Z^0$
unless specified otherwise.) In the spontaneous symmetry breaking sector,
the would--be Goldstone bosons ($\phi$'s) characterize
the broken symmetry of the
theory. These would--be Goldstone bosons become the longitudinally polarized
$W$--bosons, and the $W$--bosons therefore become massive.
Consequently, a study of the symmetry breaking sector
requires an understanding
of the interactions of these would--be Goldstone bosons. Their
interactions become strong in the TeV region
if no light resonances (Higgs bosons) exist. (In this article,
we will not consider models
with light resonances in the symmetry breaking sector.)
In the limit $E_W \gg M_W$, the S--matrix of $\WWWW$ is the same as
that of $\PPPP$, based
on the electroweak equivalence theorem.\refmark{\ETone - \ETthree}
 ($E_W$ is the energy
of the $W$--boson in the center--of--mass frame of the
$WW$ pair. $M_W$ is the
mass of the $W$--boson.) Therefore, it is important to study the
$\WWWW$ interactions in the TeV region.

In this article, we show
how to detect the $\WWWW$ signal. In section 2, we discuss the possible signals
predicted by various models. In section 3, we discuss the backgrounds.
In section 4, we discuss the characteristic differences between the event
structures of the signal and the backgrounds. In section 5,
we give recipes for detecting the signal predicted by various models.
Section 6 contains our conclusions.

\chapter{ Signal }

\REF\light{For instance, a light resonance may exist in the Standard Model,
or the Minimal Supersymmetric Model, \etc\
For a review, see J. F. Gunion, H. E. Haber, G. L. Kane and S. Dawson,
{\it The Higgs Hunter's Guide}, (Addison--Wesley, Menlo Park, 1990).}
\REF\ZZllnn{R. N. Cahn and M. S. Chanowitz, Phys. Rev. Lett. {\bf 56}, 1327
(1986);
U. Baur and E. W. N. Glover, Phys. Rev. {\bf D44}, 99 (1991), and
the references therein.}
\REF\llll{For studies on the purely leptonic mode of $WW$, see
J. Bagger, V. Barger, K. Cheung, J. Gunion, T. Han, G. Ladinsky, R. Rosenfeld
and C.--P. Yuan, in preparation,
 and the references therein.}

The event signature of the signal
is a longitudinal $W$--pair produced in the
final state. Here, we
will not discuss the detection of a light resonance
in the symmetry breaking sector as predicted by
some models.\refmark\light\ In the TeV region,
the symmetry breaking sector may either
contain a scalar-- or vector--
resonance, \etc, or no resonance at all.
For a model with a TeV scalar resonance, the most useful modes
are the $W^+W^-$ and $Z^0Z^0$ modes
 which contain large isospin--0 channel contributions.
For a model with a TeV vector resonance, the most
useful mode is the $W^\pm Z^0$ mode because it contains
a large isospin--1 channel
contribution.
If there is no resonance present in the symmetry breaking sector, all the
$WW$ modes are equally important, so the $W^\pm W^\pm$ mode is also
useful.
Before we discuss the backgrounds, we have to specify the decay mode of
the $W$'s. The branching ratio of $W^+ \ra l^+ \nu$ is 2/9 for $l^+=e^+$
or $\mu^+$, and 0.06 for $Z^0 \ra l^+l^-$.
If the signal is large enough, the
$Z^0(\ra l^+l^-)Z^0(\ra l^+l^-)$
and $Z^0(\ra l^+l^-)Z^0(\ra \nu \bar \nu)$ modes would be most
useful.\refmark{\ZZllnn}
Throughout this article, we have in mind mainly the
$\WPWM$ mode unless specified otherwise.\refmark{\llll}
We also discuss the detection of the signal in the
$\WPZ$ and $\WPWP$ modes.

\chapter{ Backgrounds }

For each decay mode of the $WW$ pair, the relevant backgrounds vary. But,
in general, the dominant background processes are the ``intrinsic''
background which also contains a $WW$ pair in the final state,
the electroweak--QCD process, $W+\,jets$,
which contains a ``fake $W$'' mimicked by two QCD jets, and QCD processes
such as $t \bar t$ production with subsequent decay to a $WW$ pair.
We now discuss these backgrounds in various $WW$ modes separately.

\section{ $\WPWM$ mode }

\REF\qqww{K. O. Mikaelian, M. A. Samuel, D. Brown, Nuovo Cim. Lett. {\bf 27}
 (1980) 211.}
\REF\ggww{C. Kao and D. A. Dicus, Phys. Rev. {\bf D43} (1991) 1555.}
\REF\wpwmew{D. A. Dicus and R. Vega, Phys. Rev. Lett. {\bf 57}, 1110 (1986);
J. F. Gunion, J. Kalinowski and A. Tofighi-Niaki, Phys. Rev. Lett. {\bf 57},
2351 (1986).}
\REF\wwjet{U. Baur, E. W. N. Glover and J. J. van der Bij, Nucl. Phys. {\bf
B318} (1989) 106;
V. Barger, T. Han, J. Ohnemus and D. Zeppenfeld,
Phys. Rev. {\bf D41} (1990) 2782, and the references therein.}
\REF\wtwoj{S. D. Ellis, R. Kleiss and W. J. Stirling, Phys. Lett. {\bf 154B}
(1985) 435; J. F. Gunion, Z. Kunszt and M. Soldate, Phys. Lett. {\bf 163B}
(1985) 389; {\bf 168B} (1986) 427 (E).}
\REF\wthreej{F. A. Berends, W. T. Giele, H. Kuijf, R. Kleiss and W. J.
Stirling, Phys. Lett. {\bf B224} (1989) 237;
V. Barger, T. Han, J. Ohnemus and D. Zeppenfeld, Phys. Rev. Lett. {\bf 62}
(1989) 1971; Phys. Rev. {\bf D40} (1989) 2888; {\bf D41} (1990) 1715 (E),
and the references therein.}
\REF\ttbaro{B. C. Combridge, Phys. Scr. {\bf 20} (1979) 5.}
\REF\ehlq{E. Eichten, I. Hinchliffe, K. Lane and C. Quigg, Rev. Mod. Phys. {\bf
56} (1984) 579; {\bf 58} (1986) 1065 (E).}
\REF\ttbarqcd{
P. Nason, S. Dawson and R. K. Ellis,
Nucl.~Phys. {\bf B303} (1988) 607; {\bf B327} (1989) 49;
W. Beenakker, H. Kuijf, W. L. van Neerven and J. Smith,
Phys.~Rev. {\bf D40} (1989) 54;
R. Meng, G. A. Schuler, J. Smith and W. L. van Neerven,
Nucl.~Phys. {\bf B339} (1990) 325.}
\REF\ttbar{R. Kauffman and C.--P. Yuan, Phys. Rev. {\bf D42} (1990) 956.}
\REF\wttbar{G. A. Ladinsky and C.--P. Yuan, Phys. Rev. {\bf D43} (1991) 789.}
\REF\ttbarjet{V. Barger, K. Cheung, T. Han and R. Phillips, Phys. Rev. {\bf
D42} (1990) 3052; D. Dicus, J. F. Gunion, L. H. Orr and R. Vega,
 Nucl. Phys. {\bf B377} (1992) 31, and the references therein.}

The intrinsic background processes for this mode are
$q \bar q \ra W^+W^-$, $g g \ra W^+W^-$
and $W^+W^-+ \, jets$.\refmark{\qqww - \wwjet}
The signature for the signal in this mode is an
isolated lepton with high transverse momentum $P_T$, and two jets which can
be reconstructed as the decay products of a $W$--boson. $W^++ \, jets$
processes\refmark{\wtwoj,\wthreej}
 can mimic the signal when the invariant mass of the two
QCD jets is around $M_W$.
Other potential background processes for this mode are the QCD processes
$ q \bar q, \, gg \ra t \bar t$, $W t \bar t$ and
$t {\bar t} + \, jet$.\refmark{\ttbaro - \ttbarjet}
For a heavy top quark, the two $W$'s
produced from the decay of $t$ and $\bar t$ can also mimic the signal.

\section{ $\WPZ$ mode }

\REF\qqwz{R. W. Brown, D. Sahdev, K. O. Mikaelian, Phys. Rev. {\bf D20} (1979)
1164.}
\REF\zttbar{ V. Barger, K. Cheung, T. Han, A. Stange and D. Zeppenfeld,
Phys. Rev. {\bf D46} (1991) 2028, and the references therein.}

The signature of the signal in this mode
is an isolated lepton with high $P_T$, a large missing transverse
energy $\ETslash$, and a two jet invariant mass around $M_Z$.
The dominant background processes for this mode are similar to those for the
$\WPWM$ mode discussed above. They are $ q_1 \bar q_2 \ra W^+ Z^0$,
$W^+Z^0 + \, jets$, $W^+ + \, jets$ and $Z t \bar t$ production
processes.\refmark{\wwjet-\wthreej,\qqwz,\zttbar}

\section { $\WPWP$ mode }

\REF\wpwpll{M. S. Chanowitz and M. Golden, Phys. Rev. Lett. {\bf 61} (1988)
1053; {\bf 63} (1989) 466 (E);
V. Barger, K. Cheung, T. Han and R. J. N. Phillips, Phys. Rev. {\bf D42} (1990)
3052, and the references therein.}
\REF\gluonex{
D. Dicus and R. Vega, Phys. Lett. {\bf 217} (1989) 194.}
\REF\wpwpew{R. Vega and D. A. Dicus, Nucl. Phys. {\bf B329} (1990) 533.}

For the purely leptonic decay
mode of $W^+W^+$,\refmark\wpwpll\ the signature is two like--sign
isolated leptons with high $P_T$ and large $\ETslash$.
There are no low--order backgrounds from quark--antiquark or gluon--gluon
fusion processes. However, other backgrounds can be important, such as
the QCD--gluon exchange
process,\refmark\gluonex\ the production of the transversely polarized
$W$--pairs
from the standard electroweak mechanism,\refmark\wpwpew\
and the $W^+t \bar t$ production from the electroweak--QCD process.

Without imposing any kinematic cuts to suppress the background processes,
the raw event rate of the signal is usually significantly
smaller than that of the backgrounds.
However, the signature of the signal can actually be distinguished from that
of the backgrounds. We shall examine the characteristic differences between the
event structures of the signal and the backgrounds in the next section.

\chapter{ How to distinguish the signal from background events }

The signature of a signal event can be
distinguished from that of background events in
many ways. We first discuss differences in the global features of the
signal and background events, then point out
some distinct kinematics of the signal events.

\section{ Global features }

\REF\wwpt{R.~N.~Cahn, S. D. Ellis, R. Kleiss, W. J. Stirling,
 Phys.\ Rev.\ {\bf D35}, 1626 (1987).}
\REF\tag{
V.~Barger, T.~Han, and R.~J.~N.~Phillips, Phys.\ Rev.\ {\bf D37}, 2005 (1988);
R.~Kleiss and W.~J.~Stirling, Phys.\ Lett.\ {\bf B200}, 193 (1988);
U.~Baur and E.~W.~N.~Glover, Nucl.\ Phys.\ {\bf B347}, 12 (1990);
D.~Dicus, J.~Gunion, L.~Orr, and R.~Vega,  Nucl. Phys. {\bf B377} (1992) 31.}

\REF\veto{
V.~Barger, K.~Cheung, T.~Han, and R.~J.~N.~Phillips,
Phys.~Rev. {\bf D42}, 3052 (1990);
D.~Dicus, J.~Gunion, and R.~Vega, Phys.~Lett.~{\bf B258},  475 (1991). }

\REF\multi{
J.~F.~Gunion, G.~L.~Kane, H.~F.-W.~Sadrozinski, A.~Seiden, A.~J.~Weinstein,
and C.--P.~Yuan, Phys. Rev. {\bf D40}, 2223 (1989).}
\REF\kane{
G.~L.~Kane and C.--P.~Yuan, Phys.~Rev. {\bf D40}, 2231 (1989). }

The signal of interest is the $WW$ pair produced from the $W$--fusion
process. The spectator quark jet that emitted the
$W$--boson in the $W$--fusion
process tends to go into the
high rapidity region. This jet typically has a high
energy, about a TeV, for $\MWW \sim 1$ TeV. ($\MWW$ is the invariant mass
of the $WW$ pair.) Therefore, one can tag this forward jet to suppress
backgrounds. \refmark{\wwpt,\tag,\llll}

Because the production of the signal is purely electroweak,
the charged particle multiplicity of the signal event is smaller than
that of a typical QCD process such as
$q \bar q \ra g W^+W^-(\ra q_1 \bar q_2)$
or $qg \ra q W^+ q_1 \bar q_2$.
Because of the small hadronic activity in the signal event, in the central
rapidity region there will be fewer hard QCD jets produced.
At the parton level, they are the two quark jets produced from the $W$--boson
decay plus soft gluon radiation.
 However, for the background process such as $t \bar t$ production,
there will be more jets produced in the central rapidity region both because
there are additional jets ($b$ and $\bar b$ jets)
from the decay of $t$ and $\bar t$ and because of the stronger hadronic
activity from QCD effects.
Therefore, one can reject events with more hard jets produced in the central
region to suppress the backgrounds. This was first suggested in
Ref.~\ttbar\ using a hadron level analysis to show that the $t \bar t$
background can be handled.

A similar trick of vetoing extra jets in the central rapidity region
when studying the pure leptonic decay mode of $W$'s
was also analyzed at the parton level.\refmark{\veto,\llll}
An equivalent way of making use of the global
difference in the hadronic activity of the events is to apply cuts on the
number of charged particles. This was first pointed out in
Refs.~\multi\ and \kane.

In the $W$--fusion process, the typical transverse momentum
of the final state $W$--pair is about $M_W$.\refmark\wwpt\
However, in the TeV region, the $P_T$ of the $W$--pair produced from the
background process, such as $q \bar q \ra gWW$, can be
$\sim$ a few hundred GeV.
Therefore, the two $W$'s (either both real or one real and one fake)
produced in the background process are less back--to--back
in the transverse plane than those of the signal.

\section{ Isolated Lepton in $W^+ \ra l^+ \nu$}

Because the background event typically has more hadronic activity in the
central rapidity region,
the lepton produced from the $W$--boson decay is
usually less isolated than that in the signal event.
 Therefore, demanding an isolated
lepton with high $P_T$ is a useful method to suppress
the backgrounds. This requirement together with
large missing transverse energy $\ETslash$ in the event
assures the presence of a $W$--boson in the event.
Also, the sign of the lepton charge can be important, as in
detecting the $\WPWP$ mode.

\section { $W \ra q_1 \bar q_2$ }

To identify the signal, we have to reconstruct the
two highest $P_T$ jets in the
central rapidity region and form the invariant mass of the
$W$--boson. It has
been shown\refmark\kane\
 that an efficient way of finding these two jets is to first
find a big cone jet with invariant mass around $M_W$, then demand that there
are two jets with smaller cone size inside this big cone jet.
Because we must measure any new activity in $\WWWW$, and
because the
$W$--boson (or ``fake $W$'') in the background event is mainly
transversely polarized,\refmark{\kane}
one must measure the fraction of longitudinally
polarized $W$--bosons in the $WW$ pair data sample and compare with that
predicted by the model of interest.

In the next section, we show how to observe the signals predicted by various
models of the
symmetry breaking sector. Some of them were studied at the hadron
level, some at the parton level. I will not reproduce those
analyses but sketch the ideas of various ``tricks'' used in detecting
$\WWWW$ interactions. The procedures discussed here are not necessarily
the ones used in the analyses performed in the literature.
If the signal event rates are large enough to observe the purely
leptonic mode, then studying the symmetry breaking sector will not be
difficult.\refmark\llll\
 Here we assume, however, that it is necessary to study the
$l^\pm + \, jets$ mode of the $WW$ pair.

\chapter{ Various Models }

\section{ A TeV Scalar Resonance }

\REF\trivial{R. Dashen and H. Neuberger, Phys. Rev. Lett. {\bf 50} (1983) 1897;
M. Lindner, Z. Phys. {\bf C31} (1986) 295, and the references therein.}
\REF\chiralone{S. Weinberg, Phys. Rev. {\bf 166} (1968) 1568;
S. Coleman, J. Wess and B. Zumino, Phys. Rev. {\bf 177} (1969) 2239;
C. Callan, S. Coleman, J. Wess and B. Zumino, Phys. Rev. {\bf 177} (1969)
2247.}
\REF\chiral{J. Gasser and H. Leutwyler, Ann. Phys. {\bf 158} (1984) 142; Nucl.
Phys. {\bf B250} (1985) 465; J. Bagger, S. Dawson and G. Valencia,
JHU-TIPAC-920009, 1992.}

In the Standard Model, the mass of the Higgs boson cannot be much larger than
$\sim 630$ GeV, otherwise the theory
would be inconsistent, based on the
triviality argument.\refmark\trivial\
(If there is any higher scale in the theory,
this number is even lower.)
However, one may consider a TeV scalar resonance
in an electroweak chiral lagrangian whose coupling
to the would--be Goldstone bosons is the same as that
in the Standard Model.\refmark{\chiralone,\chiral,\llll}
(The mass and width of the scalar resonance are two free parameters
in this model.)
Then one can ask how to detect such a TeV
scalar resonance. This study has been performed at the hadron level in
Ref.~\kane. The tricks of enhancing the ratio of signal
to background are as follows.
Let us consider the $\WPWM$ mode for this model. First of all, we trigger on a
high $P_T$ lepton. The lepton is said to be isolated if there is no more than
a certain amount of hadronic energy inside a cone
of size $\Delta R$ surrounding
the lepton. ($\Delta R=\sqrt{(\Delta \phi)^2+(\Delta \eta)^2}$,
$\phi$ is the azimuthal angle and $\eta$ is the pseudo--rapidity.)
A TeV resonance produces a $W$--boson with typical $P_T$ on the order of
$\sim 1/2$ TeV, therefore, the $P_T$ of the
lepton from the $W$--decay is on the order
of a few hundred GeV. The cut on the $P_T$ of an isolated lepton alone can
suppress a large fraction of $t \bar t$ background events because the lepton
produced from the decay of the $W$--boson typically has $P_T \sim m_t/3$, where
$m_t$ is the mass of the top quark. Besides, the lepton is also less isolated
in
the $t \bar t$ event than that in the signal event.
After selecting the events with a high $P_T$ isolated lepton, we can make use
of
the fact that the background event contains more hadronic activity than the
signal event to suppress more background. One can make a cut on the
charged particle multiplicity of the event to enhance the
signal--to--background ratio. Another way of making use of this fact is to
demand that there is only one big cone jet in the central rapidity region
of the detector. The background process typically produces more hard jets
than the signal. One can then veto the events with more than one big cone jet
in the central rapidity region. The $W^++\,jets$ and $t \bar t$ background
processes can further be suppressed by demanding that
the large cone jet has
invariant mass $\sim M_W$ and high $P_T$.
Inside this big cone jet, one can further demand two small cone
jets corresponding to the two decay quark jets of the $W$--boson.
It is essential not to bias the
information on the polarization of the $W$--boson
because discovering any new physics present requires
measuring the $W$ polarization.
 It has been shown that one
can measure the fraction of longitudinal $W$'s in the candidate $W$ samples to
distinguish various models if the event rate is not too small.\refmark\kane\
It is important to separate signal from background by general topological
aspects of events rather than by cuts.
One of the techniques which would not bias the polarization of the $W$--boson
is counting the charged particle multiplicity inside the big cone jet. A real
$W$--boson decays into a color singlet state of $q \bar q$
with the same multiplicity regardless of its energy.
 Therefore the
charged particle multiplicity of these two jets is less than that of a pair
of non--singlet QCD jets (either quark or gluon jets) which form a big cone jet
and exhibit more hadronic activity. This provides an
additional tool in suppressing the background.

\REF\misset{F. E. Paige, BNL-46828, 1991.}

Up to this point, we have only discussed the event structure in the central
rapidity region. As discussed in the previous section, in the large rapidity
region the signal event tends to have an energetic forward jet. It has been
shown that tagging one such forward jet can further suppress the background
at very little cost to the signal event rate.\refmark\llll\
Furthermore, with rapidity coverage down to 5, one can have a good
measurement on $\ETslash$.\refmark\misset\ Because
the typical $\ETslash$ due to the neutrino
from the $W$--boson decay is on the order of a few hundred GeV, the
mis--measurement of neutrino transverse momentum
due to the underlying hadronic
activity is negligible.\refmark\misset\
 Knowing $\ETslash$ and the momentum of the
lepton, one
can determine the longitudinal momentum of the
neutrino up to a two--fold solution
by constraining the invariant mass of the lepton and neutrino to
be $M_W$.\refmark\kane\
{}From the invariant mass of $l,\,\nu,\,q_1$, and $\bar q_2$, one can
reconstruct
$M_{WW}$ and distinguish
different signals from the background.
As pointed out earlier, the best way to detect new physics is to
measure the fraction ($f_L$) of longitudinal $W$'s in the event sample. A
specific model will give a distinct distribution of $f_L$ as a function
of $\MWW$. Some examples were shown in Ref.~\kane.

\section{ A TeV vector Resonance }

\REF\techni{E. Farhi and L. Susskind, Phys. Rept. {\bf 74} (1981) 277;
A review of the current status of technicolor models is given by B. Holdom,
{\it Model Building in Technicolor}, lectures given at the Nagoya Spring
School,
1991, DPNU-91-27.}

An example of this type of resonance is a techni--rho in the techni--color
model.\refmark\techni\
 What we have in mind is a vector resonance in the electroweak chiral
lagrangian. The mass and width of the vector resonance are two free parameters
in this model. Because this resonance gives a large contribution in the
isospin--1 channel,
the useful mode in which to look for such a resonance is the $W^\pm Z^0$ mode.
If the signal event rate is large enough, the resonance can be observed by
the purely leptonic decay mode
$W^+(\ra l^+\nu)Z^0(\ra l^+l^-)$ because all the leptons in this mode have
$P_T\sim$ few hundred GeV and the leptons are isolated. If the $\WPZ$ mode is
necessary for the signal to be observed, the strategies discussed in the
previous subsection for the $\WPWM$ mode can be applied in this case as well.
Needless to say, the invariant mass of two jets peaks around $M_Z$ not $M_W$.
It could be very valuable to improve techniques to separate $W( \ra jj)$
from $Z(\ra jj)$ using mass resolution and jet charge measurement as
pioneered in the JADE and ALPHA detectors.

\section{ No Resonance }

\REF\nlf{S. G. Naculich and C.--P. Yuan, Phys. Lett. {\bf B293} (1992) 405.}

If there is no resonance, then all the $WW$ modes should be measured to study
the dynamic symmetry breaking sector.\refmark\nlf\
 For instance, one may use a chiral
lagrangian model with the
lowest order two--derivative term to describe the $WW$
interactions. This model is known as the low energy theorem model. The methods
of detecting the signal from this model using the $\WPWM$ mode in the TeV
region
were presented in Ref.~\kane.
The tricks of observing this signal are identical to those discussed in the
previous subsections.
Similar tricks can be applied to observe the signal using the
$\WPZ$ mode. It has also been argued that one can
study the purely leptonic mode $\WPWP$ in
the multi--TeV region  to probe the symmetry breaking sector of
the low energy theorem model if the signal is large enough.
The dominant backgrounds
for this mode are $W^+ t \bar t$, QCD--gluon
exchange and standard electroweak processes.
To trigger this signal event, one demands two like--sign
charged leptons with high $P_T(\sim $ few hundred GeV). One further requires
these leptons to be isolated and vetos events with additional high $P_T$ jets
in
the central rapidity region.
Since there are two missing neutrinos in this event, the signal event
has large $\ETslash$, and it is difficult to reconstruct the $W$--boson and
measure its polarization. Therefore, in the
absence of a ``bump'' structure in any
distribution, we have to know the background event rate
well to study the symmetry breaking sector using this mode
unless the signal rate is large.
Similarly, measuring the charged or total
particle multiplicity of the event and tagging
a forward jet can further improve the ratio of signal to background.

Particularly for this case of no resonance, where the signal is not
large, it is very important to avoid cuts that reduce the signal or
bias a polarization measurement. There is a further technique, proposed
in Ref.~\kane, that will probably have to be used to study the
no--resonance case, and can improve our ability to study
the examples discussed above.
This technique takes advantage of the fact that the Standard Model
is well tested, and will be much better tested in the
TeV region by the time the study of $W_LW_L$
interactions is under way.
Thus, every event of a real or fake $W_LW_L$ interaction is
either a SM one or new physics. The real SM ones
(from $q \bar q , \, gg \ra WW$, $Wjj$, $t \bar t$, \etc)
can all be calculated and independently measured (in
other modes or other regions of kinematic variables).
Thus one can make global cuts such as requiring a high energy
spectator jet and low total event multiplicity,
as discussed above, and then examine all remaining candidate events
to see if they are consistent with SM processes
or if they suggest new physics, in particular new sources of longitudinal
$W$'s. In principle, only one new number needs to be measured: the fraction
of $W_LW_L$ events compared to the total number of all $WW$ events
including real and fake $W$'s. This can be done by the
usual approach of a maximum likelihood analysis, or
probably even better by the emerging neural network techniques,\Ref\neural{
B. Denby, Fermilab-92/121, 1992.}
for which it appears to be ideally suited.

Ultimately, recognizing that {\it in the TeV region}
every event is either well understood Standard Model physics or new
physics will be the most powerful approach to discovering
any deviations from the perturbative Standard Model predictions.

\section{ Beam Pipe $W$'s }

\REF\CG{R. S. Chivukula and M. Golden, Phys. Lett. {\bf B267} (1991) 233.}
\REF\inelast{S. G. Naculich and C.--P. Yuan, Phys. Lett. {\bf B293} (1992)
395.}

So far, we have only discussed signal events with high
$P_T$ $W$--bosons produced in the central rapidity region. If there are many
inelastic channels open in the $WW$ scattering
process,\refmark{\CG,\inelast,\nlf}
then based on the optical
theorem, the imaginary part of the forward elastic scattering amplitude is
related to the total cross section, and therefore will not be
small.\refmark\nlf\ This
implies that it is possible for the final state $W$'s to
predominantly go down
the beam pipe when produced from
$W^+W^- \ra W^+W^-$ elastic scattering.
Assuming this to be the case, it is
important to know how to detect such beam pipe $W$'s in the TeV region.

\REF\beam{G. Kane, S. Mrenna, S. G. Naculich and C.--P. Yuan, in preparation.}

The typical transverse momentum of the $W \ra f_1 f_2 $ decay
products is about $M_W/2$.
For $M_{WW}  > 2 M_W$,
the typical opening angle between the decay products
of one of the $W$'s is about  ${4M_W / M_{WW}}$.
Therefore, the absolute value of the rapidity
of the decay products is likely
to be within the range  2.5 to  4 for $M_{WW} \sim 1$ TeV.
With appropriate effort they can be detected (perhaps not in
every detector, but certainly in some detector eventually).
To suppress the backgrounds, one can veto
events with any jets or leptons in the central rapidity region,
$|\eta| \le 2.5$.
Another signature of the
signal event is the appearance of an energetic quark jet,
the quark recoiling after emitting one of the interacting $W$'s,
with rapidity in the range 3 to 5.
One can thus further suppress QCD and electroweak backgrounds
by tagging one forward (or backward) jet.
The background due to $W$'s emitted
in a minimum bias event can also be suppressed,
because, unlike the longitudinal $W$'s of the signal,
these $W$'s tend to be transversely polarized.
As a result, one of their decay products tends
to be boosted more than the other,
and is likely to be lost down the beam pipe, say, $|\eta| > 5$.
Combining these techniques,
we conclude that it may be feasible to detect
longitudinal $W$ scattering
even in models in which $W$'s tend to be
scattered predominantly along the beam pipe direction.\refmark\beam\

\section{ Conclusions}

\REF\velt{H. Veltman and M. Veltman, Acta. Phys. Polon. {\bf B22}, 669 (1991).}

We have discussed how to experimentally study the symmetry
breaking sector by observing
$\WWWW$ interactions in the TeV region,
emphasizing general features of the event structure in the
signal and background events. Various tricks
to enhance the ratio of signal to background
were presented. We showed how to detect longitudinal $W$--bosons
either in the central rapidity region of the detector or in
the beam pipe direction. We conclude that
if there is no light resonance present
then it is possible to study the
symmetry breaking sector in the TeV region
even when the $W_LW_L$ scattering is not resonant, as may be the most
likely outcome.\refmark\velt\
 However, to ensure a complete study of the symmetry breaking
sector, the beam
pipe $W$'s also need to be measured if no signal events
are found in the central rapidity region.

\REF\sdcgem{Some studies beyond the parton level were listed in
Ref.~\kane. For more references, see, for example,
Solenoidal Detector Collaboration, {\it
Technical Design Report}, SDC-92-201, 1992, and
GEM Collaboration, {\it Letter of Intent}, SSCL-SR-1184, 1991.}

Most of the proposals discussed here have been examined at the parton level
but not in detector simulations.\refmark\sdcgem\
 They have been demonstrated to be
promising techniques, but we cannot be sure they will work
until the detector simulations are carried out by experimentalists.
Fortunately, there will be plenty of time to do those studies before the
data is available.

\ack

I would like to thank
J. Bagger, V. Barger, E. Berger, K. Cheung, S. Dawson, J. Gunion, T. Han,
G. Kane, R. Kauffman,
G. Ladinsky, S. Mrenna, S. G. Naculich, F. E. Paige, R. Rosenfeld,
J.~L.~Rosner,
H. F.-W. Sadrozinski, A. Seiden, S. Willenbrock, A. J. Weinstein
and Y.--P. Yao for collaboration.
To them and
U. Baur,  M. S. Chanowitz,  M. Einhorn, S. D. Ellis,
Gordon Feldman, M. K. Gaillard,
E. W. N. Glover, H. Haber, C. Kao, S. Meshkov,
F. Olness, L. H. Orr, W. W. Repko, M. E. Peskin, E. Poppitz,
 C. R. Schmidt, W. J. Stirling,
Wu--Ki Tung, G. Valencia, H. Veltman, M. Veltman, R. Vega
and  E. Wang, I am grateful for discussions.
This work was funded in part by the TNRLC grant \#RGFY9240.

\refout

\end